\documentclass[prb,showpacs,showkeys,reprint,onecolumn]{revtex4-1}

\usepackage{amssymb}
\usepackage{amsmath}
\usepackage{amsfonts}
\usepackage{amsbsy}
\usepackage{graphicx}
\usepackage{float}
\usepackage{color}

\newcommand{\pp}{{\mbox{\boldmath $p$}}}
\newcommand{\qq}{{\mbox{\boldmath $q$}}}
\newcommand{\kk}{{\mbox{\boldmath $k$}}}
\newcommand{\xx}{{\mbox{\boldmath $x$}}}
\newcommand{\xxb}{{\mbox{\boldmath $\overline{x}$}}}
\newcommand{\pint}{\int^{\!-}_{\!\!-}\!\!\!\!\!\int}
\newcommand{\bq}{\begin{eqnarray}}
\newcommand{\eq}{\end{eqnarray}}
\newcommand{\tpi}{\tilde{\pi}}

\newcommand{\bwt}{\begin{widetext}}
\newcommand{\ewt}{\end{widetext}}

\begin{document}
\title{Low temperature acoustic polaron localization}  

\author{Riccardo Fantoni}
\email{rfantoni@ts.infn.it}
\affiliation{National Institute for Theoretical Physics (NITheP) and
Institute of Theoretical Physics,  University of
Stellenbosch, Stellenbosch 7600, South Africa} 

\date{\today}

\begin{abstract}
We calculate the low temperature properties of an acoustic polaron in
three dimensions in thermal equilibrium at a given temperature using 
a specialized path integral Monte Carlo method.
In particular we find numerical evidence that the chosen Hamiltonian
for the acoustic  
polaron describes a phase transition from a localized state to an 
unlocalized state for the electron as the phonons-electron coupling
constant decreases.
The phase transition manifests itself with a jump discontinuity in
the potential energy as a function of the coupling
constant. In the weak coupling regime the electron is in an extended
state whereas in the strong coupling regime it is found in a
self-trapped state.
\end{abstract}

\pacs{71.38.-k,71.38.Fp,71.38.Ht}
\keywords{Acoustic Polaron, Path Integral Monte Carlo Simulation,
  Localization} 

\maketitle
\section{Introduction}
\label{sec:introduction}

An electron in an ionic crystal polarizes the lattice in its
neighborhood. An electron moving with its accompanying
distortion of the lattice has sometimes been called a ``polaron''
\cite{Gerlach1991,Devresee2009}. 
Since 1933 Landau addresses the possibility whether an electron can be
self-trapped (ST) in a deformable lattice
\cite{Landau1933,Landau1946,Landau1948}. This fundamental problem in
solid state physics has been intensively studied for an optical
polaron in an ionic crystal
\cite{Frohlich1950,Frohlich1954}
\cite{Feynman1955,Peeters1982,Mitra1987,Mason1986}. 
Bogoliubov approached the polaron strong coupling limit with one of
his canonical transformations. Feynman used his path integral
formalism and a variational principle to develop an all coupling
approximation for the polaron ground state \cite{Feynman}. Its
extension to finite temperatures appeared first 
by Osaka \cite{Osaka1959,Osaka1965}, and more recently by Castrigiano
{\it et al.} \cite{Castrigliano1983,Castrigliano1984,Khandekar1986}. 
Recently the polaron problem has gained new interest in explaining the
properties of the high 
$T_c$ superconductors \cite{Sheng}. The polaron problem has been
also studied to describe impurities of lithium atoms in Bose-Einstein
ultracold quantum gases condensate of sodium atoms 
\cite{Tempere2009}. In this context evidence for a transition
between free and ST polarons is found whereas for the
solid state optical polaron no ST state has been found 
yet \cite{Feynman1955,Peeters1982,Mitra1987}. The
Bogoliubov dispersion at low $k$ is similar to that of acoustic phonons 
rather than to that of optical phonons.  

The acoustic modes of lattice vibration are known to be responsible
for the appearance of the ST state
\cite{Toyozawa1961,Kuper1963,Gerlach1991}. Contrary to 
the optical mode which interacts with the electron through Coulombic
force and is dispersionless, the acoustic phonons have a linear
dispersion coupled to the electron through a short range potential
which is believed to play a crucial role in forming the ST
state \cite{Peteers1985}. Acoustic modes have also been widely
studied \cite{Gerlach1991}. Sumi and Toyozawa 
generalized the optical polaron model by including a coupling to the
acoustic modes \cite{sumi73}. Using Feynman's variational approach,
they found that the electron is ST with a very large effective mass
and small radius as the acoustic coupling exceeds a critical
value. Emin and Holstein also 
reached a similar conclusion within a scaling theory \cite{Emin1976} in
which the Gaussian trial wave function is essentially identical to the
harmonic trial action used in the Feynman's variational approach in
the adiabatic limit \cite{Fisher1986}. 

The ST state distinguishes itself from an extended state (ES) where
the polaron has lower mass and a bigger radius. A polaronic phase
transition separates the two states with a breaking of translational
symmetry in the ST one \cite{Gerlach1991}. The variational approach is
unable to clearly assess the existence of the phase transition
\cite{Gerlach1991}. Nevertheless Gerlach and L\"owen
\cite{Gerlach1991} concluded that no phase transition exists in a
large class of polarons. The three dimensional acoustic polaron is not
included in this class but Fisher {\it et al.} \cite{Fisher1986} argued
that its ground state is delocalized.  

In this work we employ for the first time a particular path integral (PI)
Monte Carlo (MC) method \cite{ceperley95,Pierleoni2001} to the
continuous, highly non-local,  
acoustic polaron problem at low temperature which is valid at all
values of the coupling 
strength and solves the problem exactly. Our
method differs from 
previously employed methods
\cite{Alexandrou1990,Alexandrou1992b,Crutz1981,Takahashi1983,Wang1998,Kornilovitch1997,Kornilovitch2007}
since hinges on the L\'evy construction and the multilevel Metropolis
method \cite{ceperley95}. We
calculate the potential energy and show that like the effective mass
it usefully signals the transition between the ES
and the ST state. Our results indicate the existence
of the phase transition.

The paper is organized as follows: in Sec. \ref{sec:model} we describe
the acoustic polaron mathematical model. In Sec. \ref{sec:observables}
we describe the observables we are interested in. Sec. \ref{sec:pi}
contains the description of the numerical scheme used to solve the
path integral. In Sec. \ref{sec:results} we report our numerical
results. And Sec. \ref{sec:conclusions} is for final remarks. 

\section{The model}
\label{sec:model}
The acoustic polaron can be described by the following quasi-continuous
model \cite{Frohlich1954,sumi73},
\bq\nonumber
\hat{H}=\frac{\hat{\pp}^2}{2m}+\sum_{\kk}\hbar\omega_k
\hat{a}_{\kk}^{\dagger}\hat{a}_{\kk} + \sum_{\kk} \left(i \Gamma_k
\hat{a}_{\kk} e^{i\kk \hat{\xx}} + \mbox{H.c.}\right).
\eq
Here $\hat{\xx}$ and $\hat{\pp}$ are the electron coordinate and momentum
operators respectively and $\hat{a}_\kk$ is the annihilation operator 
of the acoustic phonon with wave vector $\kk$. The electron coordinate
$\xx$ is a continuous variable, while the phonons wave vector $\kk$ is
restricted by the Debye cut-off $k_o$. The first term in the 
Hamiltonian is the kinetic energy of the electron, the second term the
energy of the phonons and the third term the coupling energy between
the electron and the phonons with an {\it interaction vertex}
$\Gamma_k=\hbar u k_o (S/\rho_0)^{1/2}(k/k_o)^{1/2}$ where $S$ is the
coupling constant between the electron and the phonons and  
$\rho_0$ the number density of unit cells in the crystal ($\rho_0=(4\pi/3)
(k_o/2\pi)^3$ in the Debye approximation). 
The acoustic phonons have a dispersion relation $\omega_k=uk$, $u$ 
being the sound velocity.

Using the path integral representation (see Ref. \cite{Feynman}
section 8.3), the phonon part in the
Hamiltonian can be exactly integrated owing to its quadratic form in
phonon coordinates, and one can write the partition
function for a polaron in thermal equilibrium at an absolute
temperature $T$ ($\beta=1/k_BT$, with $k_B$ Boltzmann constant) as
follows, 
\bq
Z=\int d\xx \pint_{\xx=\xx(0)}^{\xx=\xx(\hbar\beta)} e^{
-\frac{1}{\hbar}{\cal S}[\xx(t),\dot{\xx}(t),t]} {\cal D}\xx(t), 
\eq 
where the action ${\cal S}$ is given by \cite{feynman65},\footnote{
This is an approximation as $e^{-\beta\omega_k}$ is neglected. The
complete form is obtained by replacing $e^{-\omega_k|t-s|}$ by
$e^{-\omega_k|t-s|}/(1-e^{-\beta\omega_k})+
e^{\omega_k|t-s|}e^{-\beta\omega_k}/(1-e^{-\beta\omega_k})$.
But remember that $\beta$ is large.}
\bq \label{action}
{\cal S}&=&\frac{m}{2}\int_0^{\hbar\beta}\dot{\xx}^2(t)dt-\frac{1}{2\hbar}
\int_0^{\hbar\beta}dt\int_0^{\hbar\beta}ds \int\frac{d\kk}{(2\pi)^3} 
\Gamma_k^2 e^{i\kk\cdot(\xx(t)-\xx(s))-\omega_k|t-s|}\\
&=&{\cal S}_f+{\cal U}.
\eq
Here ${\cal S}_f$ is the {\it free particle action}, and ${\cal U}$ the 
{\it inter-action} and we denoted with a dot a time derivative as
usual. Setting $\hbar=m=uk_o=k_B=1$ the inter-action becomes, 
\bq \label{inter-action}
{\cal U}=
\int_0^{\beta}dt\int_0^{\beta}ds \,V_{eff}(|\xx(t)-\xx(s)|,|t-s|),
\eq 
with the electron moving subject to an effective {\it retarded potential},
\bq  \label{veff}
V_{eff}&=&-\frac{S}{2I_D}\int_{q\leq 1}d\qq\, q e^{i\sqrt{\frac{2}{\gamma}}
\qq\cdot(\xx(t)-\xx(s))-q|t-s|},
\eq
where $\qq=\kk/k_o$, $I_D=\int_{q\leq 1} d\qq=4\pi/3$, and we have 
introduced a non-adiabatic parameter $\gamma$ defined as the ratio of the 
average phonon energy, $\hbar u k_o$, to the electron band-width, 
$(\hbar k_o)^2/2m$. This parameter is of order of $10^{-2}$ in typical
ionic crystals with broad band ($\sim$eV) so that the ST state is
well-defined \cite{sumi73}. In our simulation we took
$\gamma=0.02$. One can expect two kinds of polarons: Electrons in
alkali halides and silver halides are nearly ES while holes in alkali
halides are in the ST state \cite{Kanzig1955}. The hole is ES in AgBr
\cite{Hanson1960} and ST in AgCl \cite{Hohne1968}. The most dramatic
observation of the abrupt change of exciton from ES to ST states was
made on mixed crystals AgBr$_{1-x}$Cl$_{x}$ \cite{Kanzaki1971}

\section{The observables}
\label{sec:observables}

The free energy (the ground state energy, $E$, in the large $\beta$
limit) of the polaron is 
$F=-(\partial Z/\partial\beta)/Z=\langle {\cal K}+{\cal P} \rangle$, 
where the first term is the kinetic energy contribution,
${\cal K}$, and the second term is the potential energy contribution,
${\cal P}$. We have,
\bq
F&=&=\frac{1}{Z}\int d\xx
\pint e^{-{\cal S}}\frac{\partial {\cal S}}{\partial \beta}{\cal D}\xx
=\left\langle\frac{\partial {\cal S}}{\partial \beta}\right\rangle.
\eq 
Scaling the Euclidean time $t=\beta t^\prime$ and $s=\beta s^\prime$
in Eq. (\ref{action}), deriving ${\cal U}$ with respect to $\beta$ and
undoing the scaling in the end we obtain for the potential, 
\bq \nonumber
{\cal P}&=&-\frac{3S}{2\beta}\int_0^\beta dt\int_0^\beta ds\int_0^1 dq\,q^3
\frac{\sin\left(\sqrt{\frac{2}{\gamma}}q|\xx(t)-\xx(s)|\right)}
{\sqrt{\frac{2}{\gamma}}q|\xx(t)-\xx(s)|}
e^{-q|t-s|}(2-q|t-s|).
\eq
Taking the derivative with respect to $\beta$ of the action after
having scaled both the time as before and the coordinate
$\xx=\sqrt{\beta}\xx^\prime$ and undoing the scaling in the end we
obtain for the kinetic energy,
\bq
{\cal K}=-\frac{3S}{4\beta}\int_0^\beta dt\int_0^\beta ds\int_0^1 dq\,q^3
\left[\cos\left(\sqrt{\frac{2}{\gamma}}q|\xx(t)-\xx(s)|\right)-
\frac{\sin\left(\sqrt{\frac{2}{\gamma}}q|\xx(t)-\xx(s)|\right)}
{\sqrt{\frac{2}{\gamma}}q|\xx(t)-\xx(s)|}\right]e^{-q|t-s|}.
\eq

In the following we will be concerned with a numerical determination
of the potential energy.

\section{Path integral Monte Carlo}
\label{sec:pi}

To calculate the PIMC, we first choose a 
subset of all paths. To do this, we divide the independent variable,
Euclidean time, into {\it steps} of width $\tau = \beta/M$.
This gives us a set of {\it times}, $t_k=k\tau$ spaced a distance $\tau$
apart between $0$ and $\beta$ with $k=0,1,2,\ldots,M$.
At each time $t_k$ we select the special point $\xx_k=\xx(t_k)$, the $k$th
{\it time slice}. We construct a path by connecting all points so selected 
by straight lines. It is possible to define a sum over all paths constructed
in this manner by taking a multiple integral over all values of $\xx_k$
for $k=1,2,\ldots,M-1$ where $\xx_0=\xx_a$ and $\xx_M=\xx_b$ are the two 
fixed ends.
The simplest discretized expression for the action can then be written 
as follows,
\bq \label{s-discr}
{\cal S}=\sum_{k=1}^M \frac{(\xx_{k-1}-\xx_k)^2}{2\tau}+\tau^2\sum_{i=1}^M
\sum_{j=1}^M V(t_i,t_j),
\eq 
where $V(t_i,t_j)=V_{eff}(|\xx_i-\xx_j|,|i-j|)$ is a symmetric  
two variables function. In our simulation we tabulated 
this function taking $|\xx_i-\xx_j|=0,0.1,0.2,\ldots,10$ and $|i-j|=
0,1,\ldots,M$.
The total configuration space to be integrated over is made of elements
$s=\{\xx_0,\xx_1,$ $\ldots,\xx_M\}$ where $\xx_k$ are the path time slices 
subject to the periodic boundary condition $\xx_M=\xx_0$.
In order to compute the potential energy $P=\langle {\cal P}\rangle$
in the simulation we wish to sample these elements from the probability 
distribution, $\pi(s)=e^{-{\cal S}}/Z$, 
where the partition function $Z$ normalizes the function $\pi$ in this 
space.

In our simulation we chose to use the bisection method, a particular 
multilevel MC method
\cite{ceperley86,ceperley89,ceperley95}, with correlated
sampling. The {\it transition probability} for the first level is
chosen as   
$T_1\propto \exp[(\xx_{i+m/2}-\xxb)^2/2\sigma^2(m/2)]$
where $m=2^l$, $l$ being the number of levels,
$\xxb=(\xx_i+\xx_{i+m})/2$ and $\sigma(t_0/\tau)=\sqrt{\langle
  [\xx(t)-(\xx(t+t_0)+\xx(t-t_0))/2]^2\rangle}$ (for the first levels
these deviations are smaller than the free particle 
standard deviations used in the L\'{e}vy construction \cite{levy39}
$\sigma_f(\ell)=\sqrt{\ell\tau/2}$ with $\ell_k=m/2^k$ in the
$k$th level. Much smaller in the ST state.). And so on for the
other levels: 
$s_2=\{\xx_{i+m/4},\xx_{i+3m/4}\}$, $\ldots$ ,
$s_l=\{\xx_{i+1},\xx_{i+2},\ldots,\xx_{i+m-1}\}$. And
$s_0=\{\xx_0,\ldots,\xx_i,\xx_{i+m},\ldots,\xx_{M-1}\}$ where $i$ is
chosen randomly.     
Calling $\tpi(s)=e^{-{\cal U}}/Z$, the {\it level inter-action} is
$\tpi_k(s_0,\ldots,s_k)=\int ds_{k+1}\ldots ds_{l}\,\tpi(s)$.
For the $k$th level inter-action we thus chose the following 
expression,
\bq \label{laction1}
\tpi_k\propto \exp\left[-(\tau\ell_k)^2\sum_{i=1}^{[M/\ell_k]}
\sum_{j=1}^{[M/\ell_k]}V(i\ell_k\tau,j\ell_k\tau)\right].
\eq 
In the last level $\ell_l=1$ and the level
inter-action $\tpi_l$ reduces to the exact inter-action $\tpi$.
The acceptance probability for the first level will then be,
$A_1=\min\left[1,\frac{P_1(s)}{P_1(s^\prime)}\frac{\tpi_1(s^\prime)
\tpi_{0}(s)}{\tpi_1(s)\tpi_{0}(s^\prime)}\right]$
with $P_1\propto\exp\{-(\xx_{i+m/2}-\xxb)^2[
1/\sigma^2(m/2)-1/\sigma_f^2(m/2)]/2\}$.
The initial path was chosen with all time slices set to $\vec{0}$.
During the simulation we maintain the acceptance ratios in
$[0.15,0.65]$ by decreasing (or 
increasing) the number of levels in the multilevel algorithm as the
acceptance ratios becomes too low (or too high). We will call Monte
Carlo step (MCS) an attempted move. 

\section{Results}
\label{sec:results}

\begin{figure}[H]
\begin{center}
\includegraphics[width=12cm]{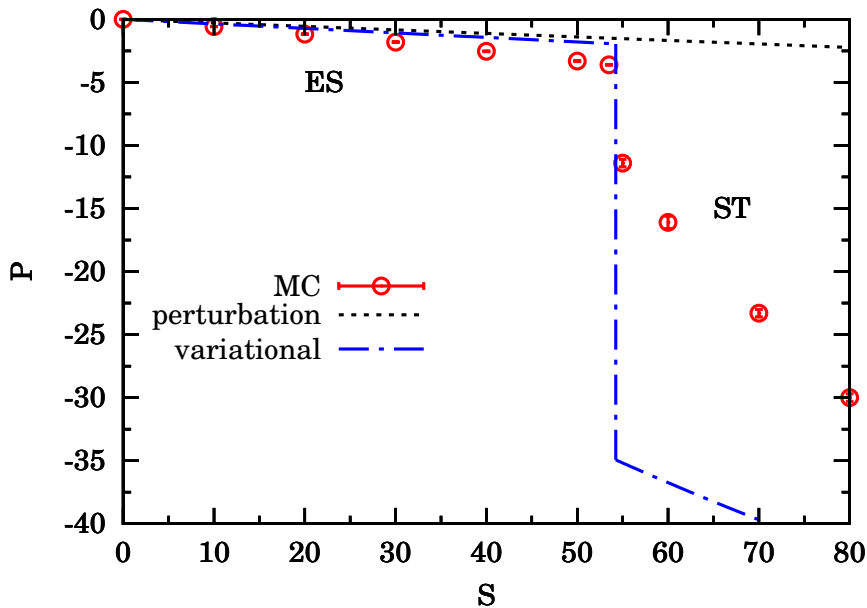}\\
\end{center}  
\caption{Shows the behavior of the potential energy $P$ as a function
  of the coupling constant $S$. The points are the MC results (see 
  Tab. \ref{table}), the dashed line is the second order 
  perturbation theory result (perturbation) and the dot-dashed line is
  the variational approach from Ref. \cite{sumi73} (variational)
  in the weak and strong coupling regimes.}
\label{fig:s-p}
\end{figure}

\begin{figure}[H]
\begin{center}
\includegraphics[width=12cm]{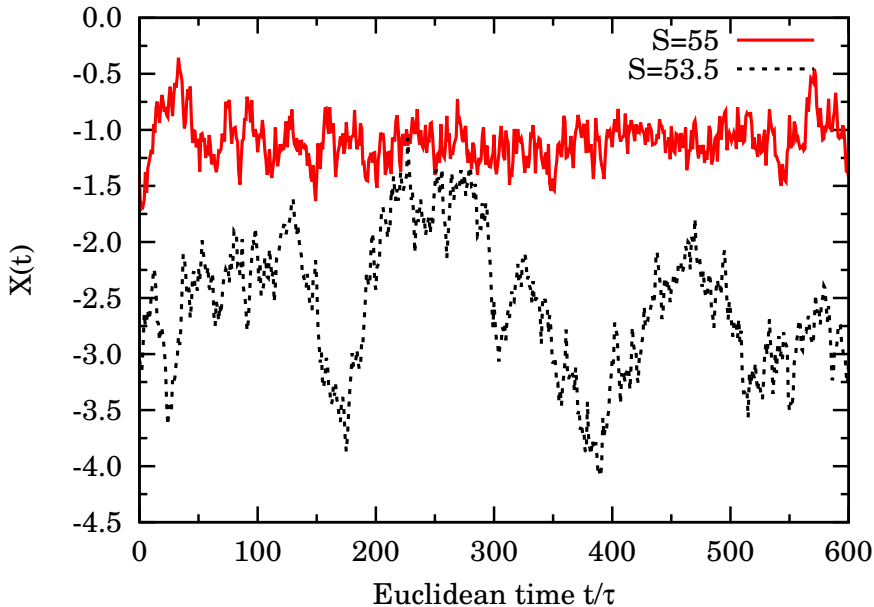}
\end{center}  
\caption{The polaron closed path $\xx(t)=(X(t),Y(t),Z(t))$ as a
  function of Euclidean time $t$ in units of $\tau$.} 
\label{path}
\end{figure}

We simulated the acoustic polaron fixing the adiabatic
coupling constant $\gamma=0.02$ and the inverse temperature $\beta=15$.
Such temperature is found to be well suited to extract close to ground
state properties of the polaron \cite{Wang1998}. 
For a given coupling constant $S$ we computed the potential energy
$P$ extrapolating (with a linear $\chi$ square
fit) to the continuum 
time limit, $\tau\to 0$, three points corresponding to time-steps
choosen in the interval $\tau\in[1/100,1/30]$.
In Fig. \ref{fig:s-p} and Tab. \ref{table} we show the results for the
potential energy as a function of the coupling strength. 
\begin{table}[hbt]
\caption{MC results for $P$ as a function of $S$ at $\beta=15$ and
  $\gamma=0.02$ displayed in Fig. \ref{fig:s-p}. The runs where made of
  $5\times 10^5$ MCS (with $5\times 10^4$ MCS for the equilibration)
  for the ES states and $5\times 10^6$ MCS (with $5\times 10^5$ MCS
  for the equilibration) for the ST states.} 
\label{table}
{\scriptsize
\begin{center}
\begin{tabular}{||c||c||}
\hline
$S$ & $P$ \\
\hline
\hline
10&  -0.573(8)\\
20&  -1.17(2)\\
30&  -1.804(3)\\
40&  -2.53(3)\\
50&  -3.31(4)\\
53.5&-3.61(1)\\
55&  -11.4(3)\\
60&  -16.1(5)\\
70&  -23.3(3)\\
80&  -30.0(3)\\ 
\hline
\end{tabular}
\end{center}
}
\end{table}
It is clear the transition
between two different regimes which correspond to the so called ES
and ST states for the weak and strong coupling region respectively. We
found that paths related to ES and ST are characteristically
distinguishable. Two typical paths for the ES and ST regimes involved
in Fig. \ref{fig:s-p} are illustrated in Fig. \ref{path}. The path in
ES state changes smoothly on a large time scale, whereas the path in ST
state do so abruptly on a small time scale with a much smaller
amplitude which is an indication that the polaron hardly moves. The
local fluctuations of the $\xx(t)$ and of the potential energy ${\cal 
P}$(MCS) have an auto-correlation function which decay much more slowly
in the ES state than in the ST one. Moreover the ES simulations are
more time consuming than the ST ones. 

Concerning the critical property of the transition between the ES and
ST states our numerical results are in favor of the presence of a
discontinuity in the potential energy. Even if there is no trace of a
translational symmetry breaking as shown by the ST path in 
Fig. \ref{path} where the initial path was $\xx(t)=\vec{0}$ for all
$t$. With the increase of $\beta$, the values for the potential energy
$P$ increase in the weak coupling regime but
decrease in the strong coupling region. 
From second order perturbation theory (see Ref. \cite{Feynman} section
8.2) follows that the energy shift $E(\gamma,S)$ is given by
$-3S\gamma[1/2-\gamma+\gamma^2\ln(1+1/\gamma)]$ from which one
extracts the potential energy shift by taking
$P(\gamma,S)=\gamma dE(\gamma,S)/d\gamma$. From the
Feynman variational 
approach of Ref. \cite{sumi73} follows that in the weak regime the
energy shift is $-3S\gamma[1/2-\gamma+\gamma\ln(1+1/\gamma)]$ and in
the strong coupling regime $-S+3\sqrt{S/5\gamma}$.
Note that since $S$ and $\tau$ appear in the combination $S\tau^2$ in
${\cal U}$ (and $S\tau$ in ${\cal P}$) the same phase transition from
an ES to a ST state will be observed increasing the temperature. With
the same Hamiltonian we are able to describe two very different
behaviors of the acoustic polaron as the temperature changes.

\section{Conclusions} 
\label{sec:conclusions}

In conclusion we used for the first time a specialized PIMC 
method to study the low temperature behavior of an acoustic
polaron. At an inverse temperature $\beta=15$, close to the ground
state of the polaron, and at a non-adiabatic parameter $\gamma =0.02$,
typical of ionic crystals, we found numerical evidence for a phase
transition between an ES state in the weak coupling regime and a
ST one in the strong coupling regime at a value of the
phonons-electron coupling constant $S^\star=54.3(7)$ in good agreement with
the prediction of Ref. \cite{sumi73}, $S^\star\sim 1/\gamma$, and the MC
simulations of Ref. \cite{Wang1998}.

To understand the motion of an electron in a deformable lattice we
have to consider the fact that the interaction
of the electron with the acoustic phonons
induces a well barrier proportional to the coupling constant, but it
decays due to the retarded property. In the weak coupling region, the
electron can easily tunnel through the barrier so that it almost
freely moves in the lattice. One can regard the tunneling of the
electron as an indication of the motion of the polaron. In this case,
a few phonons are involved and the acoustic polaron has a small mass
which has a similar magnitude to the mass of the electron and a large
radius. In the 
strong coupling region, the well barrier becomes sufficiently deep and
the electron is temporarily bounded in the polaron and cannot tunnel
through the barrier until it gains enough energy from the
phonons. Much more phonons are involved in the polaron
and the mass of the polaron becomes much greater than that of the
electron with a small radius. In this argument the specific form of
the interaction vertex is of fundamental importance.

We used for the first time a PIMC with the bisection method and
correlated sampling as an unbiased numerical mean to probe the
low temperature properties of the acoustic polaron. This is 
an independent route to the Feynman's variational approach and 
proved to give reliable results on the existence of the ST
state for the electron in a deformable lattice as conceived by
Landau. However, the self-trapping we observe in our numerical
analysis is not a complete localization of the electron within the
polarization cloud of the phonons. The electron path still undergoes
small, nearly uncorrelated, fluctuations in Euclidean time in this ST
state. Our numerical results support the presence of a discontinuity
in the potential energy as a function of the coupling constant and
this would be an indication of the existence of a phase transition
between the ES and the ST states even if we found no trace of a
translational symmetry breaking. Moreover The discontinuity of the
ES/ST transition, if it exists, may depend on the cutoff parameter as
pointed out by e.g. Ref. \cite{Peeters1982}. In the cold 
atoms context the role of this parameter is important and the
existence of a discontinuous transition is more questionable than in the
solid state polaron case \cite{Tempere2009}. The present study reports
a single value study of the acoustic polaron case thus restricting the
generality of the conclusions. 

In a truly localized state the polaron should not diffuse at all
(strong localization) or at least should attain a subdiffusive
behavior (weak localization). But these properties can be checked by
looking at the real time dynamics of the system and cannot be
checked by the Monte Carlo methods as that used in this work which
deals with polaron properties in imaginary time. Clearly our
numerical MC results support the claim of a dicontinuity but do not
give any proof of a localization transition.
 
A possible further study could involve the dynamic properties
associated with the two different types of motion and bipolarons for
short range interacting systems.

\begin{acknowledgments}
R.F. would like to thank David Ceperley
for suggesting the problem and for his guidance during the preparation
of the work.
\end{acknowledgments}

%

\end{document}